\documentstyle[aas2pp4]{article}

\slugcomment{Submitted to The Astrophysical Journal}
\lefthead{Semer\'ak, de Felice, \& Karas}
\righthead{A method to determine the parameters\ldots}

\begin{document}

\title{A method to determine the parameters of black holes in
       AGNs\\
       and galactic X-ray sources with periodic modulation of
       variability}
\author{Old\v{r}ich~Semer\'ak,\altaffilmark{1}
        Fernando~de~Felice\altaffilmark{2,3} and
        Vladim\'{\i}r Karas\altaffilmark{4}}
\altaffiltext{1}{Department of Theoretical Physics,
                 Faculty of Mathematics and Physics,
                 Charles University,
                 V~Hole\v{s}ovi\v{c}k\'ach~2,
                 CZ-180\,00 Praha~8,
                 Czech Republic;
                 semerak@mbox.troja.mff.cuni.cz}
\altaffiltext{2}{Department of Physics ``G.~Galilei'',
                 University of Padova,
                 Via Marzolo~8,
                 I-35131 Padova,
                 Italy}
\altaffiltext{3}{INFN, Sezione di Padova;
                 defelice@padova.infn.it}
\altaffiltext{4}{Astronomical Institute,
                 Faculty of Mathematics and Physics,
                 Charles University,
                 V~Hole\v{s}ovi\v{c}k\'ach~2,
                 CZ-180\,00 Praha~8,
                 Czech Republic;
                 karas@mbox.troja.mff.cuni.cz}

\begin{abstract}
We propose a simple and unambiguous way to deduce the parameters of
black holes which may reside in AGNs and some types of X-ray binaries.
The black-hole mass and angular momentum are determined in physical
units. The method is applicable to the sources with periodic components
of variability, provided one can assume the following:
(i)~Variability is due to a star or a stellar-mass compact
    object orbiting the central black hole and passing
    periodically through an equatorial accretion disk
    (variability time-scale is given by the orbital period).
(ii)~The star orbits almost freely, deviation of its trajectory due to
     passages through the disk being very weak (secular);
     the effect of the star on the disk, on the other
     hand, is strong enough to yield observable photometric and
     spectroscopic features.
(iii)~The gravitational field within the nucleus is that of the
      (Kerr) black hole, the star and the disk contribute
      negligibly.
\end{abstract}


\section{Introduction}

There is an ever-growing theoretical and observational evidence that the
dark masses present in some galactic nuclei and X-ray binaries are black
holes (Rees 1998). Several ways have been suggested of how to
deduce the properties of these black holes. The first rough estimates of
their mass were based on energy considerations and limits implied by the
shortest variability timescale (e.g.\ Begelman, Blandford, \& Rees
1984). More precise methods became possible with modern observational
techniques like HST, VLBI and X-ray satellites. The aim of the present
paper is to propose and further discuss a method to determine the
parameters of a black hole in a system with two periodic components in
the observed signal which are due to orbital motion of a stellar object
and its passages through the accretion disk of the hole. We start with a
brief summary of the subject.

In galactic nuclei, the existence and size of the dark central mass are
deduced from the curve of the central surface brightness of the nuclei
and from the spatial distribution and dynamics of surrounding gas and
stars (see Kormendy \& Richstone 1995 for a review and references). In
particular, the standard model for active galactic nuclei (AGNs) with a
black hole and an accretion disk (e.g.\ Rees 1984; Blandford \& Rees
1992) has been further supported when the nuclear gas in several active
galaxies was found to be in a gravity-dominated nearly circular motion
in a disk (e.g.\ Jaffe et al.\ 1993; Gallimore, Baum, \& O'Dea 1997).
Rotation curves of the observed nuclear gas disks indicate a central
compact dark mass
of some (2.0--3.5)$\times$$10^{9}M_{\odot}$ in M\,87 (NGC 4486)
   (Merritt \& Oh 1997; Macchetto et al.\ 1997),
of about 4.9$\times$$10^{8}M_{\odot}$ in NGC 4261
   (Ferrarese, Ford, \& Jaffe 1996), and
of 3.6$\times$$10^{7}M_{\odot}$ in NGC 4258
   (Miyoshi et al.\ 1995; Maoz 1995).

The observed excess of quasars at high redshifts has stimulated a
search for black holes in quiescent galaxies, since they
may have been both the engine and then the residue of a former
activity (Haehnelt \& Rees 1993). In this case the most reliable
present data come from the observations of stellar motion (van
den Bosch \& de Zeeuw 1996). Velocity profiles of several stellar
nuclear disks offer strong evidence for black holes ---
of 2.6$\times$$10^{6}M_{\odot}$ in our own Galaxy
   (Genzel et al.\ 1997; cf.\ Mezger, Duschl, \& Zylka 1996),
of 2$\times$$10^{9}M_{\odot}$ in NGC 3115
   (Kormendy et al.\ 1996a),
of 1$\times$$10^{9}M_{\odot}$ in NGC 4594 (the Sombrero galaxy)
   (Kormendy et al.\ 1996b),
of 3$\times$$10^{6}M_{\odot}$ in M\,32
   (van den Marel et al.\ 1997), and probably also
of 6$\times$$10^{8}M_{\odot}$ in NGC 4486B
   (Kormendy et al.\ 1997).

However, current optical and radio studies still probe only regions
above $10^{4}\div 10^{5}$ gravitational radii of the putative holes
which is by far not enough to resolve any imprints of the relativistic
effects due to the collapsed centre, and in particular to deduce the
rate of its rotation.

Information from within the region of a few tens of gravitational radii
have been brought by X-ray satellite observations of a sample of Seyfert
1 galaxies (Nandra et al.\ 1997 and references therein; see also Fabian
1997 for a survey, and Tanaka et al.\ 1995 for the best-quality
measurement from MCG-6-30-15). These observations yielded the profiles
of the fluorescent iron K$\alpha$ emission line which is likely to be
induced by illumination of the very inner parts of an accretion disk by
a source off the disk plane (e.g.\ Matt, Perola, Piro, \& Stella 1992;
Petrucci \& Henri 1997). Indeed, the observed broad profile skewed to
lower energies is best explained in terms of the Doppler and
gravitational redshifts (Fabian et al.\ 1995); the exact origin of the
illuminating source remains unclear. Evidence for smallness of the
emitting regions also comes from the resolved rapid variability of the
K$\alpha$ lines (Nandra et al.\ 1997, and references therein). Note that
recently Bao et al.\ (1997) reminded another unique (and measurable)
feature of black hole sources involving accretion disk: the
energy-dependent variability of polarization of their X-rays,
originally discussed by Stark \& Connors (1977).

A way to estimate parameters of the central black hole is strongly
related to the source variability. It is assumed, in fact, that a time
delay between variations of the emission line strength of the AGN and
that of the continuum can be due to the time travel between the central
source and the surrounding line-emitting gas. Radial distance and
velocity of propagation of the disturbances impose limits on the
black-hole mass (Blandford \& McKee 1982; Krolik et al.\ 1991), but
the resulting estimates inherit uncertainties of the underlying assumptions.
Alternatively, it has been proposed to deduce the masses of the central
nuclear bodies from temporal changes of the observed emission lines
(Stella 1990), while rotational parameters from the position, intensity,
width (Hameury, Marck, \& Pelat 1994; Martocchia \& Matt 1996), and
profile of the lines (Bromley, Chen, \& Miller 1997; Dabrowski et al.\
1997; Reynolds \& Begelman 1997; Rybicky \& Bromley 1998; cf.\ also Bao,
Hadrava, \& {\O}stgaard 1994 and 1996). The X-ray photometric
light-curve profile from a hot spot orbiting not far from the horizon
also reflects rotation of the black hole (Asaoka 1989; Karas 1996).

Similar research has been focused on galactic X-ray binaries where a
central engine was also proposed, namely an accretion disk fed by
overflow from a secondary star. Here the evidence for a black hole
involves non-stellar appearance (``invisibility'') of one of the
components, the presence of X-radiation, variability, spectral features
(mainly the presence of relatively strong ultrasoft component and of
hard X-ray tail), and of course a large lower limit on mass of the dark
component ($\gtrsim3M_{\odot}$), deduced from the orbital parameters of the
(stellar) companion (namely from the mass function of the system). (See
the surveys in Lewin, van Paradijs, \& van den Heuvel 1995; the reviews
of black-hole binaries with a thorough list of references have also been
given by Barret, McClintock, \& Grindlay 1996; Charles 1997. Later
results are due to Beekman et al.\ 1996 and 1997; Orosz \& Bailyn 1997.)

It has been proposed recently to infer the parameters of the dark
component from observable behaviour of the disk. An observable
modulation of the X-ray emission may be caused by disk vibrations whose
lowest frequency depends on the mass and rotation of the hole (Nowak et
al.\ 1997 and references therein). Zhang et al.\ (1997) proposed, also
on the basis of the standard thin accretion disk model, how the angular
momentum (spin) of the black hole in an X-ray binary could be inferred
from the strength of an ultrasoft X-component. Very recently Cui, Zhang,
\& Chen (1998) have proposed ``that certain types of quasi-periodic
oscillations (QPO) observed in the light-curves of black-hole binaries
are produced by X-ray modulation at the precession frequency of
accretion disks, because of relativistic dragging of inertial frames
around spinning black holes''. Given the mass of the hole, they are able
to derive its spin by comparing the computed disk precession frequency
with that of the observed quasi-periodic oscillations. This mechanism
requires a warped accretion disk, the assumption that clearly calls for
further investigation (Markovi\'c \& Lamb 1998). Scheme for the source
variability discussed in the present paper is different from that which
has been proposed for QPOs in the above-mentioned papers. Here, an
orbiting companion of the black hole is involved --- the assumption
which restricts variability time-scales relevant for the model.

Narayan, McClintock, \& Yi (1996), and Narayan, Garcia, \& McClintock
(1997) argued that the black-hole X-ray binaries could also be
recognized according to a larger variation in luminosity between their
bright and faint states than is expected in the sources with neutron
stars. This way of identifying black holes follows from the low
radiative efficiency of ad\-vec\-t\-ion-do\-mi\-na\-ted accretion flows
which could occur around black holes (with greater ``sucking-in'' power
and no solid surface) rather than around other compact objects. Various
under-luminous accretion-powered astrophysical systems, mainly quiescent
transient X-ray sources and low-luminosity galactic nuclei, could be
interpreted as black holes with advection-dominated accretion disks (see
Lasota \& Abramowicz 1997 for a survey); Reynolds et al.\ (1996)
suggested that advection-dominated mode of the final stages of accretion
could also account for the ``quiescence'' of the black hole in M\,87.
The lack of a hard surface of a black hole also plays crucial role in an
alternative explanation by King, Kolb, \& Szuszkiewicz (1997) of X-ray
transients which are dynamical black-hole candidates, namely in terms of
a weaker (stabilizing) irradiation of the disk by the central accreting
source.

In the present paper we propose a method applicable, under certain
assumptions, to the black-hole sources with periodic components of
variability. It provides, unambiguously, {\em{}both the mass and the
specific angular momentum of the black hole in physical units}.
We start from the model of
possible periodic variability of black-hole sources which considers a
thin accretion disk lying in the hole's equatorial plane, and an object
(a ``star'') which intersects the disk periodically while orbiting about
the centre.\footnote
{See e.g.\ Krolik et al.\ 1991; Mineshige, Ouchi \& Nishimori 1994;
 Ipser 1994; Zakharov (1994); Bao, Hadrava, \& {\O}stgaard (1994);
 Kanetake, Takeuti, \& Fukue (1995), and references therein for
 alternative explanations of periodic or quasi-periodic variability.}
Such a system is characterized by two angular frequencies --- that of
azimuthal revolution of the star and that of its latitudinal oscillation
about the equatorial plane. Both frequencies are in principle measurable
at infinity, the azimuthal one from spectrophotometry, while the
latitudinal one from photometry provided that the passages of the star
produce a strong enough modulation of the source. Karas \&
Vokrouhlick\'y (1994) illustrated, by Fourier analysis of simulated
photometric data, how the respective two peaks can be recognized in the
power spectrum. This model was cultivated notably in connection with the
NGC~6814 galaxy whose putative variability, however, was later
recognized as being due to a source in our Galaxy. Other
possible targets are discussed in the current literature. For
example, optical outbursts in the blazar OJ~287 have recently been
modelled in terms of a black-hole binary system by Villata et al.\
(1998). (The source exhibits several time-scales: feature-less
short-term variability, 12-yr cycle, and, possibly, a 60-yr cycle.)
These authors, however, presume both components to be of comparable
masses while in our calculation frequencies are determined under
assumption that the secondary is much less massive than the primary
(cf. also Sundelius et al.\ 1997). As another example, a 16-hr
periodicity in the X-ray signal from the Seyfert galaxy IRAS 18325-5926
was described by Iwasawa et al.\ (1998).

Let us suppose that the disk has only a weak dynamical influence on the
star and that stellar tides are negligible (the star is assumed much
smaller than the typical curvature radius of the field around) as well
as gravitational radiation and self-gravity both of the
disk and of the star itself. Then the worldline of the star is very
close to a geodesic in a pure gravitational field of the central
rotating black hole. This is described by the Kerr metric which reads
(Misner, Thorne \& Wheeler 1973, p.\ 878), in Boyer-Lindquist spheroidal
coordinates ($t,r,\theta,\phi$), in geometrized units (in which $c=G=1$,
$c$ being the speed of light in vacuum and $G$ the gravitational
constant) and with the ($-$+++) signature of the metric tensor
($g_{\mu\nu}$),
\begin{eqnarray} \label{metric}
  {\rm d}s^{2} & = &
  -\frac{\Delta\Sigma}{\cal A}\,{\rm d}t^{2}
  +\frac{{\cal A}}{\Sigma}\,\sin^{2}\theta\;
   ({\rm d}\phi-\omega_{\rm K}{\rm d}t)^{2} \nonumber \\ & &
  +\frac{\Sigma}{\Delta}\,{\rm d}r^{2}
  +\Sigma{\rm d}\theta^{2},
\end{eqnarray}
where $M$ and $a$ denote mass and specific rotational angular
momentum of the source and
\begin{equation}
  \Delta=r^{2}-2Mr+a^{2}, \;\;\;
  \Sigma=r^{2}+a^{2}\cos^{2}\theta,
\end{equation}
\begin{equation}
  {\cal A}=(r^{2}+a^{2})^{2}-\Delta a^{2}\sin^{2}\theta, \;\;\;
  \omega_{\rm K}=2Mar/{\cal A}.
\end{equation}

It has been demonstrated (Syer, Clarke, \& Rees 1991; Vokrouhlick\'y \&
Karas 1993, 1998) that under the above-described circumstances the stellar
orbit undergoes three secular changes: a decrease of the semi-major axis
(the star spirals towards the centre due to the loss of energy in
collisions with the disk), circularization (the orbit becomes spherical,
$r={\rm const}$), and a decrease of the amplitude of precession of the
orbital plane about the equatorial plane of the centre (the orbit
gradually declines into the disk plane). Since the time scale for
circularization is typically found shorter than, or of the same order
as, the time scale necessary to drag the orbit into the disk, one can
expect that at late stages of evolution of the hole-disk-star system the
star follows a nearly equatorial spherical geodesic. This stage is also
the one in which the star stays for a relatively long time.

In the next section we discuss relevant properties of a precessing orbit
in the Kerr spacetime. Then, in Sect. \ref{eqset}, we derive simple
relations which appear to be appropriate for practical study of the
relevant objects. Equations further simplify if the object orbits not
very close to the black hole, as discussed in Sect. \ref{simple}. Our
method employs also the observed emission-line profiles which were
vastly studied in recent literature; we summarize the relevant formulae
and results in Appendix~A.

\section{The Orbiting Star as a ``Pharaoh's Fan''}
\label{Fan}

For a general spherical geodesic in the Kerr spacetime, the
frequencies of the azimuthal and latitudinal motion are given by
rather complicated formulas involving elliptic integrals (Wilkins
1972; Karas \& Vokrouhlick\'y 1994). However, the formulas
simplify considerably in the case of a nearly equatorial
geodesic. The azimuthal angular velocity
$\omega={\rm d}\phi/{\rm d}t$ (with respect to an observer at
rest at radial infinity) can be approximated by that of the
equatorial circular geodesic (Bardeen, Press, \& Teukolsky 1972),
\begin{equation}  \label{omega}
  \omega_{\pm}=(a+1/y_{\pm})^{-1},
\end{equation}
where
$y_{\pm}=y(\omega_{\pm})=\pm\sqrt{M/r^{3}}$ are the
corresponding values of the ``reduced frequency'', in general
defined by $y=\omega/(1-a\omega)$ (de Felice 1994), and the
upper/lower sign corresponds to prograde/retrograde
orbit.\footnote
{We allow for both ($\pm$) cases for completeness, but solely the
 prograde trajectory can be considered: the accretion disk is
 more likely to be corotating with the central black hole and it
 has been shown that the interaction with the disk makes
 also the star corotate eventually.}
The angular frequency $|\Omega|$ of (small) harmonic
latitudinal oscillations about the equatorial plane is given, for
a spherical orbit with {\em general} (but of course steady)
radial component of acceleration, by
\begin{equation}  \label{Omega}
 \Omega^{2}=
 \left(u^{t}/r\right)^{2}\left\{\Delta\omega^{2}+
       2y_{\pm}^{2}\left[a-(r^{2}+a^{2})\,\omega\right]^{2}\right\},
\end{equation}
where
\begin{eqnarray}  \label{ut}
  \left(u^{t}\right)^{-2}
  &=&-g_{tt}-2g_{t\phi}\omega-g_{\phi\phi}\omega^{2} \nonumber \\
  &=&1-2Mr\Sigma^{-1}(1-a\omega\sin^{2}\theta)^{2} \nonumber \\
  & &-(r^{2}+a^{2})\,\omega^{2}\sin^{2}\theta
\end{eqnarray}
is the time component of the corresponding four-velocity, and
$\omega={\rm{}const}$. The expression (\ref{Omega}) was obtained
by de Felice \& Usseglio-Tomasset (1996), in studying the
behaviour of a gedanken apparatus baptized ``Pharaoh's fan'' (de
Felice \& Usseglio-Tomasset 1992), as a result of the analysis of
the equation of relative deviation; for a simpler derivation by
perturbation of the equatorial circular orbit, see Semer\'ak \&
de Felice (1997).

We will briefly describe the ``Pharaoh's fan'' because the
equations describing its behaviour are relevant also for the
present work. The device consists of a monopole test particle in
a non-friction short narrow pipe which constrains the particle's
motion in two spatial dimensions but leaves it free in the
remaining direction. It was considered with the purpose to
provide a space traveller with a measurement which, together with
several other quasi-local experiments, yields a complete and
non-ambiguous information about spacetime and about the orbit
(Semer\'ak \& de Felice 1997). When the observer is moving on a
(generally non-Keplerian) equatorial circular trajectory in the
Kerr field and has the pipe turned to latitudinal direction,
symmetrically to the orbital (i.e.\ equatorial) plane, the centre
of the pipe is the particle's stable equilibrium position. If
released after being perturbed slightly off the equilibrium, the
particle harmonically oscillates (``the Fan waves'') about the
equatorial plane, with the proper frequency
$|\Omega|$. This cannot be made vanish by any choice
of the angular velocity $\omega$, since, in Kerr spacetime, it is
not possible to orbit freely on circular orbits at a constant
non-equatorial latitude (de Felice 1979). Except for the case
$a=0$ (the Schwarzschild spacetime around a nonrotating centre),
$\Omega^{2}$ is {\em not} equal to the square of the
proper azimuthal frequency, $(u^{t}\omega)^{2}$, as a result of
the ``dragging'' effect of mass currents existing within the
rotating source (Lense \& Thirring 1918; Wilkins 1972).

While the problem itself of the astronaut's orientability in
black-hole fields is yet of a limited practical usefulness, a
question was raised (de Felice \& Usseglio-Tomasset 1992;
Semer\'ak \& de Felice 1997) whether some kind of correspondence
could be established between the local experiments and the data
which can be measured ``at infinity''. In the case of the AGNs or
stellar-size X-ray sources, such a link is provided by the
orbiting star as discussed above since it plays the role of the
``Pharaoh's fan''; in fact, the frequency of disk crossing would
just be twice the Fan's frequency $|\Omega|$. (Since
the star is free also in the radial direction, no pipe is
necessary to constrain it.) We must also realize that what can in
principle be inferred from (photometric) observations is not the
proper frequency $|\Omega|$, but rather the value
measured by a distant observer, given by
$|\Omega_{\infty}|=|\Omega|/u^{t}$.
>From eqs. (\ref{omega})--(\ref{Omega}) the explicit expression is
\begin{equation}  \label{Omegaobs}
  \Omega_{\infty}^{2}=
  \omega_{\pm}^{2}\,(1-4ay_{\pm}+3a^{2}/r^{2}).
\end{equation}

\section{The Set of Equations for Determining
 {\protect\boldmath{$M$}}, {\protect\boldmath{$a$}} and
 {\protect\boldmath{$r$}}}
\label{eqset}

A different kind of information can be obtained from the analysis
made by Fanton, Calvani, de Felice, \& \v{C}ade\v{z} (1997, Sect.~7)
relating the quantities which are observable in the integrated
spectrum of an accretion disk to the parameters of the hole--disk
system. A stationary disk produces the well-known double-horn
line profile which is presumably modulated by the star-disk
interaction at a certain radius (see Appendix for details).
Approximating the star as an emitting point-like source on a
circular equatorial geodesic, the observed frequency shift $g$ of
each emitted photon can be written in terms of its direction
cosinus $e_{\hat\phi}$ (azimuthal component of the unit vector
along the direction of emission of a given photon, measured in
the emitter's local rest frame) as
\begin{equation}  \label{g}
 g\equiv\frac{\nu_{\rm{}obs}}{\nu_{\rm{}em}}=
 u^{t}\left[1-2M/r+y_{\pm}
 \left(a+\sqrt{\Delta}\,e_{\hat{\phi}}\right)\right],
\end{equation}
where
\begin{equation}
 \left(u^{t}\right)^{-2}=1-3M/r+2ay_{\pm}
\end{equation}
from (\ref{Omega}) and (\ref{omega}). One of the most important
attributes of a spectral line is its width; this arises from the
different frequency shifts $g$ carried by the photons which reach
the observer at infinity. Since $g$ depends on the emission
cosinus $e_{\hat{\phi}}$, the maximum broadening of the line
which could be measured as a result of an integration over one
entire orbit, is a measure of the variation $\delta g$
corresponding to the maximum range of variability of
$e_{\hat{\phi}}$ compatible with detection at infinity. If we
call the latter $\delta{}e_{\hat{\phi}}$, we have from (\ref{g})
\begin{equation}
  (\delta{g})=(\delta e_{\hat{\phi}})\,
             u^{t}\omega_{\pm}\sqrt{\Delta}
\end{equation}
and thus
\begin{equation}  \label{delta}
  \delta^{2}\equiv
  \frac{(\delta g)^{2}}{(\delta e_{\hat{\phi}})^{2}}=
  \omega_{\pm}^{2}r^{2}\,
  \frac{1-2M/r+a^{2}/r^{2}}{1-3M/r+2ay_{\pm}} \; .
\end{equation}

It is clear that $\delta{}e_{\hat{\phi}}$ can be at most 2 but, in
realistic situations, it varies significantly with the inclination angle
$\theta_{\rm{o}}$ of the black hole--disk system with respect to the
line of sight. (It also depends, though only weakly, on the rotational
parameter $a$ and the radius of emission $r$.) From a numerical
ray-tracing analysis (Fanton 1997, private communication) it is found
that $\delta e_{\hat{\phi}}$ ranges from $(\delta e_{\hat{\phi}})_{\rm
min}\simeq 0.4$ to $(\delta e_{\hat{\phi}})_{\rm max}\simeq 2$ as
$\theta_{\rm{o}}$ goes from $\simeq 0^{\circ}$ to $\simeq 90^{\circ}$,
hence one can fix as observables the extreme values of $\delta$,
$\delta_{\rm max}=2.5\,\delta g$ and $\delta_{\rm min}=0.5\,\delta g$,
(corresponding to a line of sight nearly polar in the former case and
nearly equatorial in the latter one). If the line width $\delta g$ is
measured, then formulas (\ref{omega}), (\ref{Omegaobs}) and
(\ref{delta}) provide a closed system of ordinary equations which yields
the parameters $M$, $a$ and $r$ in terms of the observable quantities
$\omega_{\pm}$, $|\Omega_{\infty}|$ and $\delta$.

These equations can be solved for $a$, $r^{2}$ and $y_{\pm}$, for
example:
\begin{eqnarray}
  a&=&\omega_{\pm}^{-1}-y_{\pm}^{-1}, \\
  r^{2}&=&\frac{3\,(1-\omega_{\pm}/y_{\pm})^{2}}
               {4\omega_{\pm}y_{\pm}+
                \Omega_{\infty}^{2}-
                5\omega_{\pm}^{2}},
\end{eqnarray}
where $y_{\pm}$ is determined by quartic equation
\begin{equation}  \label{quartic}
  34y_{\pm}^{4}-By_{\pm}^{3}+Cy_{\pm}^{2}-Dy_{\pm}+E=0
\end{equation}
with
\begin{eqnarray}
  B&=&\frac{4}{\omega_{\pm}}\,
      (\delta^{2}\omega_{\pm}^{2}+23\omega_{\pm}^{2}-
       2\Omega_{\infty}^{2}), \\
  C&=&21\delta^{2}\omega_{\pm}^{2}+76\omega_{\pm}^{2}+
      7\delta^{2}\Omega_{\infty}^{2}+
      3\Omega_{\infty}^{2}
      \nonumber \\
      & &+\Omega_{\infty}^{4}/\omega_{\pm}^{2}, \\
  D&=&\frac{2}{\omega_{\pm}}\;
      \left[18\delta^{2}\omega_{\pm}^{4}+12\omega_{\pm}^{4}+
            5\delta^{2}\omega_{\pm}^{2}
                       \Omega_{\infty}^{2}
            \right. \nonumber \\ & & \left.
            +11\omega_{\pm}^{2}\Omega_{\infty}^{2}
            -(\delta^{2}+1)\,\Omega_{\infty}^{4}
            \right], \\
  E&=&5\omega_{\pm}^{4}(4\delta^{2}+3) \nonumber \\ & &
      +(\omega_{\pm}^{2}-\Omega_{\infty}^{2})
       (\delta^{2}\Omega_{\infty}^{2}
        -7\omega_{\pm}^{2})
      -\Omega_{\infty}^{4}.
\end{eqnarray}
Evidently it would be rather cumbersome to solve eq.\
(\ref{quartic}) analytically and one better finds the result
numerically for each given set of data $\omega_{\pm}$,
$|\Omega_{\infty}|$ and $|\delta|$.

Let us note that in the Schwarzschild case, $a=0$, eqs.\
(\ref{omega}) and (\ref{Omegaobs}) reduce to
\begin{equation}  \label{oyO,Schw}
  \omega_{\pm}=y_{\pm}=\pm|\Omega_{\infty}|=
  \pm\sqrt{M/r^{3}}
\end{equation}
and eq.\ (\ref{delta}) reads
\begin{equation}
  \delta^{2}=r^{2}\omega_{\pm}^{2}\,
             \frac{1-2r^{2}\omega_{\pm}^{2}}
                  {1-3r^{2}\omega_{\pm}^{2}} \, .
\end{equation}
The physical solution of this equation is
\begin{equation}
 r^{2}=(4\omega_{\pm}^{2})^{-1}
       \left[3\delta^{2}+1
             -\sqrt{(3\delta^{2}+1)^{2}-8\delta^{2}}\right],
\end{equation}
$M$ following then from (\ref{oyO,Schw}) as
$M=r^{3}\omega_{\pm}^{2}$.

\section{A Simple Explicit Solution for
 {\protect\boldmath{$r^{2}{\gg}a^{2}$}}}
\label{simple}

\begin{table}[t]
\begin{center}
\begin{tabular}{crcccr}
\tableline
$a/M$ & \rule{0mm}{2.4ex}$r/M$ & $\omega_+$ & $\Omega_+$ &
  $\tilde{a}/\tilde{M}$ & $\tilde{r}/\tilde{M}$  \\
\tableline
0.30 &  5 & 0.0871 & 0.0827   & 0.300    &  5.8 \rule{0mm}{2.8ex}\\
     & 10 & 0.0313 & 0.0307   & 0.301    & 10.6 \\
     & 20 & 0.0111 & 0.0110   & 0.300    & 20.5 \\
     & 40 & 0.0039 & 0.0039   & 0.300    & 40.5 \\
0.60 &  5 & 0.0848 & 0.0772   & 0.616    &  5.5 \rule{0mm}{2.8ex}\\
     & 10 & 0.0310 & 0.0300   & 0.606    & 10.4 \\
     & 20 & 0.0111 & 0.0109   & 0.602    & 20.4 \\
     & 40 & 0.0039 & 0.0039   & 0.601    & 40.4 \\
0.85 &  5 & 0.0831 & 0.0735   & 0.925    &  5.3 \rule{0mm}{2.8ex}\\
     & 10 & 0.0308 & 0.0294   & 0.868    & 10.3 \\
     & 20 & 0.0110 & 0.0108   & 0.855    & 20.4 \\
     & 40 & 0.0039 & 0.0039   & 0.851    & 40.4 \\
1.00 &  5 & 0.0821 & 0.0716   & 1.200    &  5.3 \rule{0mm}{2.8ex}\\
     & 10 & 0.0306 & 0.0291   & 1.033    & 10.3 \\
     & 20 & 0.0110 & 0.0108   & 1.008    & 20.3 \\
     & 40 & 0.0039 & 0.0039   & 1.002    & 40.4 \\
\tableline
\end{tabular}
\end{center}
\tablenum{1}
\caption{Illustration of accuracy of approximative relations;
 $\tilde{a}/\tilde{M}$ and $\tilde{r}/\tilde{M}$ are estimates based on
 eqs.\ (\protect\ref{r})--(\protect\ref{a}). Notice that acceptable
 precision is reached for $r\gtrsim10M$.
 \label{tab1}}
\end{table}

Due to processes of tidal disruption close to the horizon
(Frank \& Rees 1976; Marck, Lioure, \& Bonazzola 1996; Diener et
al.\ 1997) and in general due to rather violent conditions near
the inner edge of the disk\footnote
{This is standardly considered somewhere near the marginally
 stable and marginally bound circular equatorial geodesic orbits
 which may lie, in terms of the $r$-coordinate, very
 close to the horizon ($r\simeq{M}$) when the black hole
 rotates rapidly ($a\simeq M$).}
where most of the radiation is produced, the star probably cannot orbit,
in the manner required above, on radii of the order of the centre's
gravitational radius ($\simeq{M}$). On the other hand, the radius must
not be too large since this would prevent us to recognize any
relativistic effect and even to cover a sufficient number of periods in
the observations. It is then relevant to suppose that the star orbits at
some $r\gtrsim10M$, and typically $r\sim30M$.

One can verify that some terms in eqs.\ (\ref{omega}),
(\ref{Omegaobs}) and (\ref{delta}) are then very small
when compared with the rest. Ignoring these small terms, one
arrives at a simplified set of relations which keeps an acceptable
precision of the resulting parameters without need to solve the
fourth-order equation (\ref{quartic}).
Exact equations can only then be
used to improve the estimates. It is suitable to choose some
particular approximation according to what precision one desires
at each of the quantities $M$, $a$, $r$, and also according to
precision of each of the input data
$|\omega_{\pm}|$, $|\Omega_{\infty}|$, $|\delta|$.
For instance, since $ay_{\pm}<1/30$, we can take

\begin{equation}
  \omega_{\pm}\doteq y_{\pm}(1-ay_{\pm}), \;\;\;
  y_{\pm}\doteq\omega_{\pm}(1+a\omega_{\pm})
\end{equation}
(for the determination of $M$ from the known $r$, or vice versa,
it is even sufficient to neglect $ay_{\pm}$ altogether and use
just Schwarzschild equation).
Also $\frac{a^{2}/r^{2}}{2M/r}<\frac{1}{20}$, but the other
relevant bounds are rather weak, namely
\begin{equation}
  \frac{a^{2}/r^{2}}{ay_{\pm}}=\frac{ay_{\pm}}{M/r}=
  \frac{a}{\sqrt{Mr}}<\frac{1}{5} \, ,
\end{equation}
so one must be careful when making neglections in eqs.\
(\ref{Omegaobs}) and (\ref{delta}); in particular, neglecting
the term $a^{2}/r^{2}$ in eq.\ (\ref{Omegaobs}), a very simple
formula for $a$ is obtained just by combination with eq.\
(\ref{omega}),
\begin{equation}
  a=\frac{1}{\omega_{\pm}}\,
    \frac{\omega_{\pm}^{2}-\Omega_{\infty}^{2}}
         {5\omega_{\pm}^{2}-\Omega_{\infty}^{2}} \,,
\end{equation}
but this turns out to yield inaccurate estimates.

Let us present an example of a simple and satisfactory approximation.
(We denote the approximative values by a tilde.)
Suppose the observed signal is produced by a star on a {\em{prograde}\/}
orbit ({\em{}plus\/} sign in relevant relations).
Setting the fraction in eq.\
(\ref{delta}) equal to unity, one obtains an approximative value for
the radius,

\begin{equation}  \label{r}
 \tilde{r}=\delta/\omega_{+};
\end{equation}
within the considered range of $a$ and $r$, the error is 6.5\% at
maximum (for $a=0$ and $r=10M$) but typically only 1.3\% (corresponding
to $a=0.8M$, $r=30M$). Substituting $\tilde{r}$ and $\omega_{+}$ in
place of $r$ and $y_{+}$ in eq.\ (\ref{Omegaobs}), one approximates $a$ by

\begin{equation}  \label{a}
 \tilde{a}=
 \frac{\delta}{3\omega_{+}}
 \left[2\delta-
       \sqrt{4\delta^{2}-
       3(1-\Omega_{\infty}^{2}/\omega_{+}^{2})}
 \right].
\end{equation}
$\tilde{M}$ is then given by eq.\ (\ref{omega}):
\begin{equation}
  \tilde{M}=\frac{\omega_{+}^{2}\tilde{r}^{3}}{(1+\tilde{a}\omega_{+})^{2}}
   =\frac{\delta^{3}}{\omega_{+}(1+\tilde{a}\omega_{+})^{2}} \; .
\end{equation}
The most desired information, namely the values
of $\tilde{a}/\tilde{M}$ and
$\tilde{r}/\tilde{M}$,
follow by obvious combinations from the deduced $\tilde{a}$,
$\tilde{r}$ and $\tilde{M}$.
The results of this approximation are illustrated in
Table~\ref{tab1} for several typical $a$ and $r$, together with
the respective values of the observable frequencies.

\placetable{tab1}

Before using the above formulas for some particular set of real
data, one must realize that they are written in geometrized
units. We remark that the relevant quantities in physical units
are obtained by the following conversions:
\begin{eqnarray}
  \frac{M^{\rm phys}}{M^{\rm phys}_{\odot}}&=&
     \frac{M}{1.477\times 10^{5}{\rm cm}} \, , \\
  a^{\rm phys}=ca, \;&\;&\; r^{\rm phys}=r;
\end{eqnarray}
also
\begin{equation}
  \frac{a}{M}=\frac{a^{\rm phys}}{GM^{\rm phys}/c} \,,\quad
  \frac{r}{M}=\frac{r^{\rm phys}}{GM^{\rm phys}/c^{2}} \,.
\end{equation}
To obtain frequency in physical units [Hz] (either from $\omega$ or
$\Omega$), one uses the relation
\begin{equation}
  f^{\rm phys}=\frac{\omega^{\rm phys}}{2\pi}
              =\frac{c\omega}{2\pi}
              =(4.771\times 10^{9}{\rm cm\cdot s^{-1}})\,\omega.
\end{equation}
In usual notation of relativistic astrophysics the
(geometrized) frequencies are
scaled by $M^{-1}$, as in Tab.~\ref{tab1}. Numerical values thus
obtained must be multiplied by a factor

\begin{equation}
  \frac{c}{2\pi M}=(3.231\times 10^{4}{\rm Hz})\,
                   \left(\frac{M}{M_{\odot}}\right)^{\!-1}
\end{equation}
to get the frequency in [Hz].

\section{Discussion}

In the present paper, we outlined general features of our method;
an application to particular sources will be discussed separately
(work in progress). We will conclude by brief comments on
several important points.

First, although the method described above {\em{}in principle\/} yields
{\em{}all\/} the relevant quantities, it might be
combined advantageously with independent determination of some of the
parameters. In particular, knowing $M$, one can deduce $a/M$ and $r/M$
from eqs.\ (\ref{a}) and (\ref{r}).

Observed emission-line profiles from relativistic accretion disks around
black holes have recently attracted great interest and they play a
significant role also in our method. Apart from the well-known
double-horn feature (Laor 1991), characteristics of the lines depend on
rather uncertain properties of accretion flows, e.g.\ on advection
velocity (Fukue \& Ohna 1997), limb-darkening law (Rybicki \& Bromley
1997) and shape of the disk (Pariev \& Bromley 1997). One therefore
needs to investigate a broad range of models to determine widths,
required in our model, and to reject unacceptable profiles. We present
some results in this direction in the Appendix.

Crucial point in our considerations is the (quasi)-periodic
variability of the black-hole--disk source. Periodic modulation
of the accretion flow is the most likely signature for
a stellar companion close to the central black hole
(Podsiadlowski \& Rees 1994). Each star's passage through the
accretion disk pulls some amount of gaseous material out of the
disk (Zurek, Siemiginowska, \& Colgate 1994). This material
temporarily covers the innermost region of the disk and modulates
the observed radiation (both continuum and line).\footnote
{Syer \& Clarke (1995) discussed a response of the disk to a body
 moving completely inside and determined conditions under which a
 gap can develop and survive (see also \v{S}lechta 1998).}
It can thus be anticipated that a variable signal is recognized
in the X-ray band, however, details of the mechanism remain
rather uncertain. Note that also the star may be affected by the
collisions and tidal forces considerably --- it may
lose its outer layers or even become disrupted (Frank \& Rees
1976; Marck, Lioure, \& Bonazzola 1996; Diener et al.\ 1997; Loeb
\& Ulmer 1997). A stripped stellar core is an intense source of
(ir)radiation which may survive for many orbital periods (Rees
1998). In any case, we assumed here that the star is only
{\em weakly\/} affected by the disk, i.e.\ that its orbit may
well be approximated by a nearly equatorial spherical geodesic.

Radiation from the presumed accretion disks can be modulated by their
precession, vibrations and various instabilities. It is also very well
possible that the disks are covered by numerous irregularities
(phenomenologically designated as bright spots), contributing to the
featureless short-term X-ray variability which is indeed observed
(Abramowicz et al.\ 1991; Mangalam \& Wiita 1993). Then our scheme must
anticipate that well-separated,  profound irregularities develop
on the disk surface (due to stellar passage) and survive several orbital
periods.

Although no AGN or black-hole X-ray binary with clear periodic
variability is currently known, the cases of rapidly variable
K$\alpha$ line (Iwasawa et al.\ 1996; Yaqoob et al.\ 1996), and
that of quasi-periodic AGNs (Papadakis \& Lawrence 1993;\footnote
{However, cf.\ Tagliaferri et al.\ (1996).}
Lehto \& Valtonen 1996; Stothers \& Sillanp\"a\"a 1997; Valtonen \&
Lehto 1997) and X-ray sources (Callanan et al.\ 1992; Pav\-len\-ko et al.\
1996; Belloni et al.\ 1997; Steiman-Cameron \& Scargle 1997; Iwasawa et
al.\ 1998) suggest that some periodic components are present in the
signal.
Thus we conclude with M.~J.\ Rees (1998):
{\sl{``There is a real chance that someday observers will find
evidence that an AGN is being modulated by an orbiting star,
which would act as a test particle whose orbital precession would
probe the metric in the domain where the distinctive features of
the Kerr geometry should show up clearly.''}}

\def\lb#1{{\protect\linebreak[#1]}}
\acknowledgments
O.\,S.\ thanks for support from the grants
GACR 202/\lb{2}96/\lb{2}0206 of the Grant Agency of the Czech
Republic and GAUK 230/96 of the Charles University, and
F.\,de\,F.\ for support from the Agenzia Spaziale Italiana, the
Gruppo Nazionale per la Fisica Matematica del C.N.R.\ and the
Ministero della Ricerca Scientifica e Tecnologica of Italy.
V.\,K.\ acknowledges the grant GACR 205/\lb{2}97/\lb{2}1165.
O.\,S. thanks the Director of the Department of Physics of the
University of Padova for hospitality.


\appendix

\section{Appendix}
\label{appa}
\begin{figure*}
\centering\leavevmode
\epsfxsize=18cm
\epsfbox[90 187 553 597]{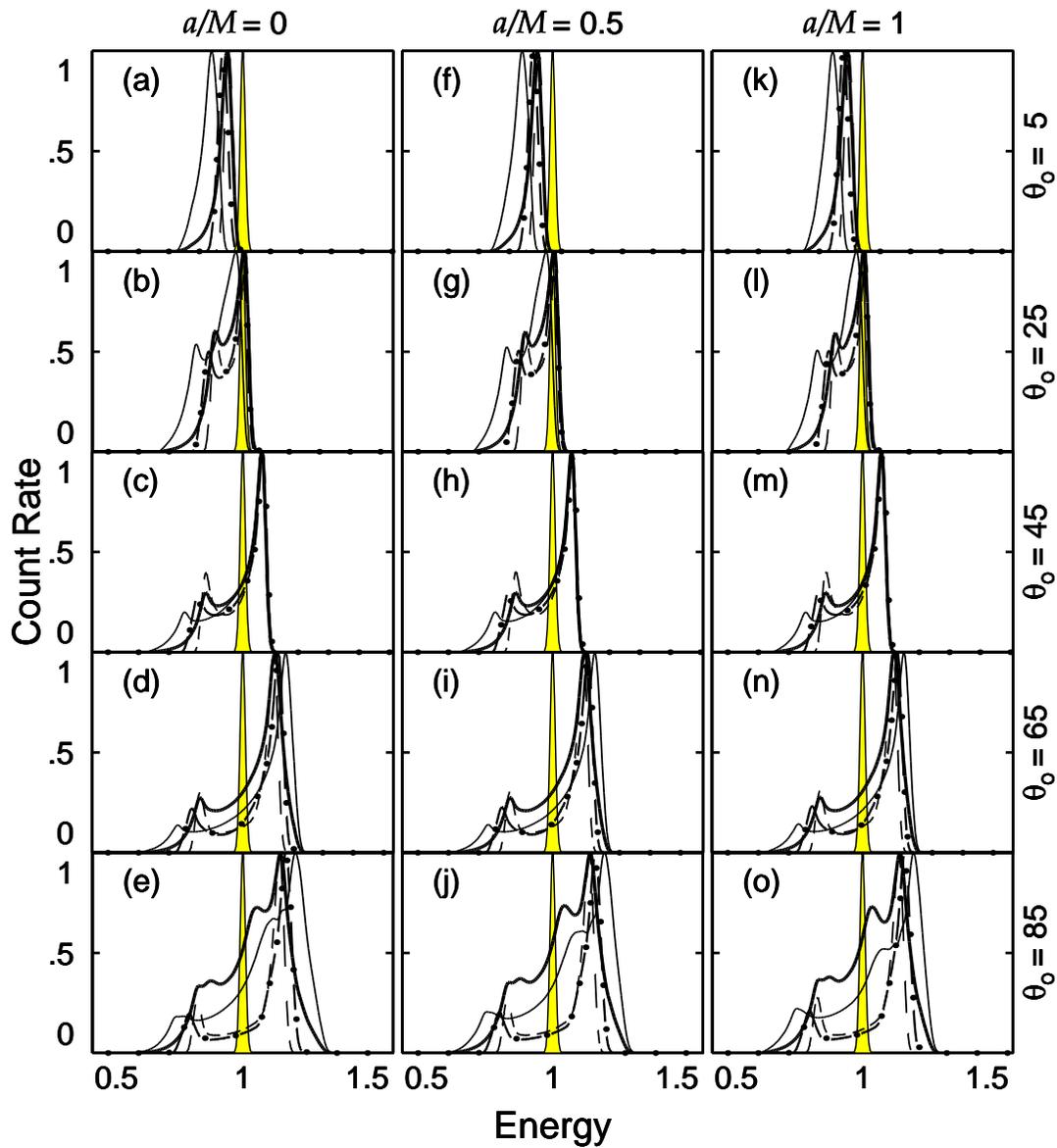}
\caption{Synthetic emission-line profiles from a thin Keplerian disk
 with locally isotropic radiation and for three values of the black-hole
 dimensionless angular momentum: $a/M=0$ (left column), $a/M=0.5$
 (center), and $a/M=1$ (right). Observer's inclination $\theta_{\rm o}$
 acquires values between $\theta_{\rm o}=0^{\rm{}o}$ (pole-on view) and
 $90^{\rm{}o}$ (edge-on). See the text for a detailed description and
 Tab.~2 for characteristics of the profiles.
 \label{fig1}}
\end{figure*}

\begin{figure*}
\centering\leavevmode
\epsfxsize=18cm
\epsfbox[90 187 553 597]{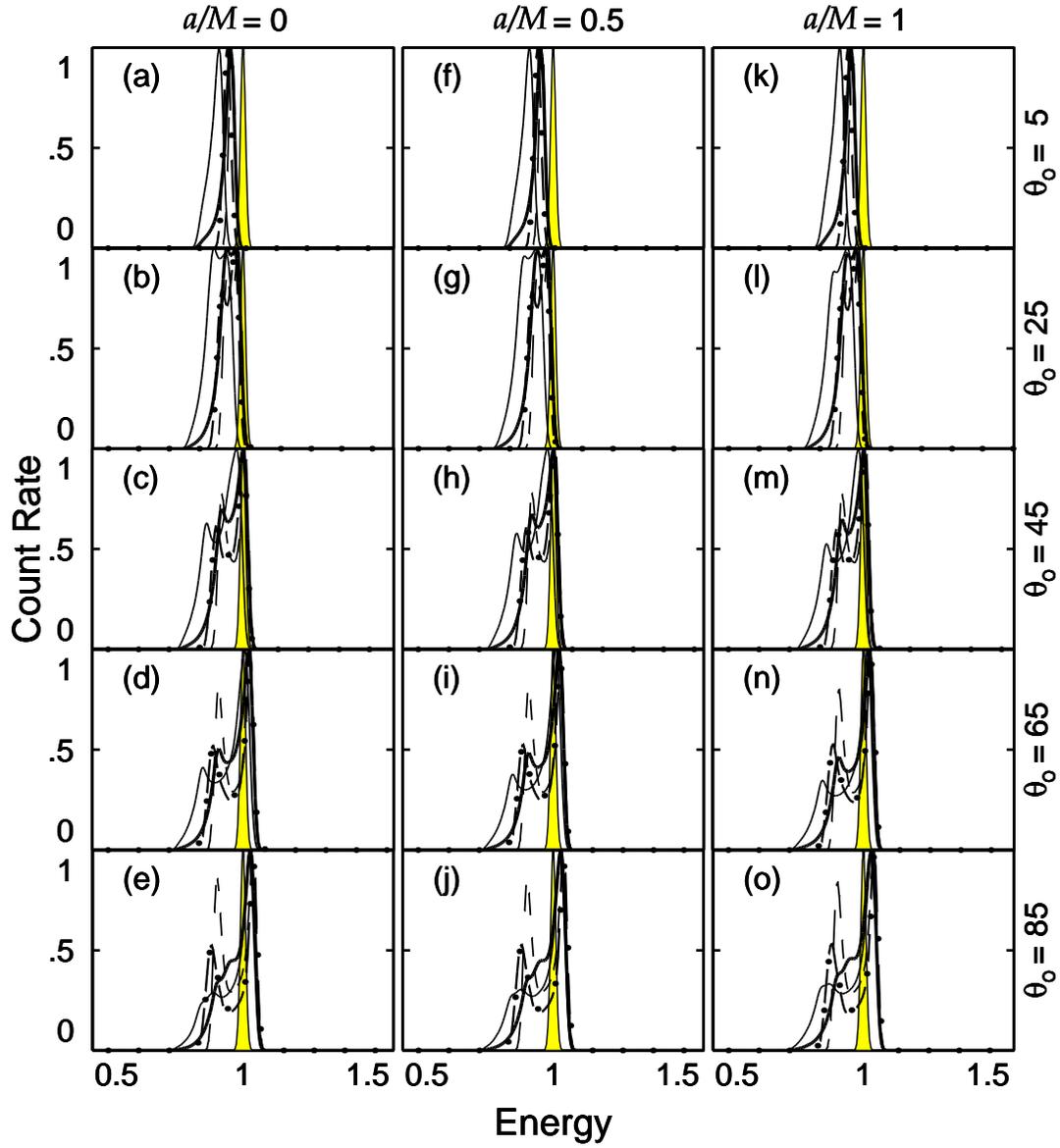}
\caption{As in Fig.~\protect\ref{fig1} but for a disk with radial
 component of the flow velocity comparable to the azimuthal component.
 \label{fig2}}
\end{figure*}

\begin{figure*}
\centering\leavevmode
\epsfxsize=18cm
\epsfbox[90 187 553 597]{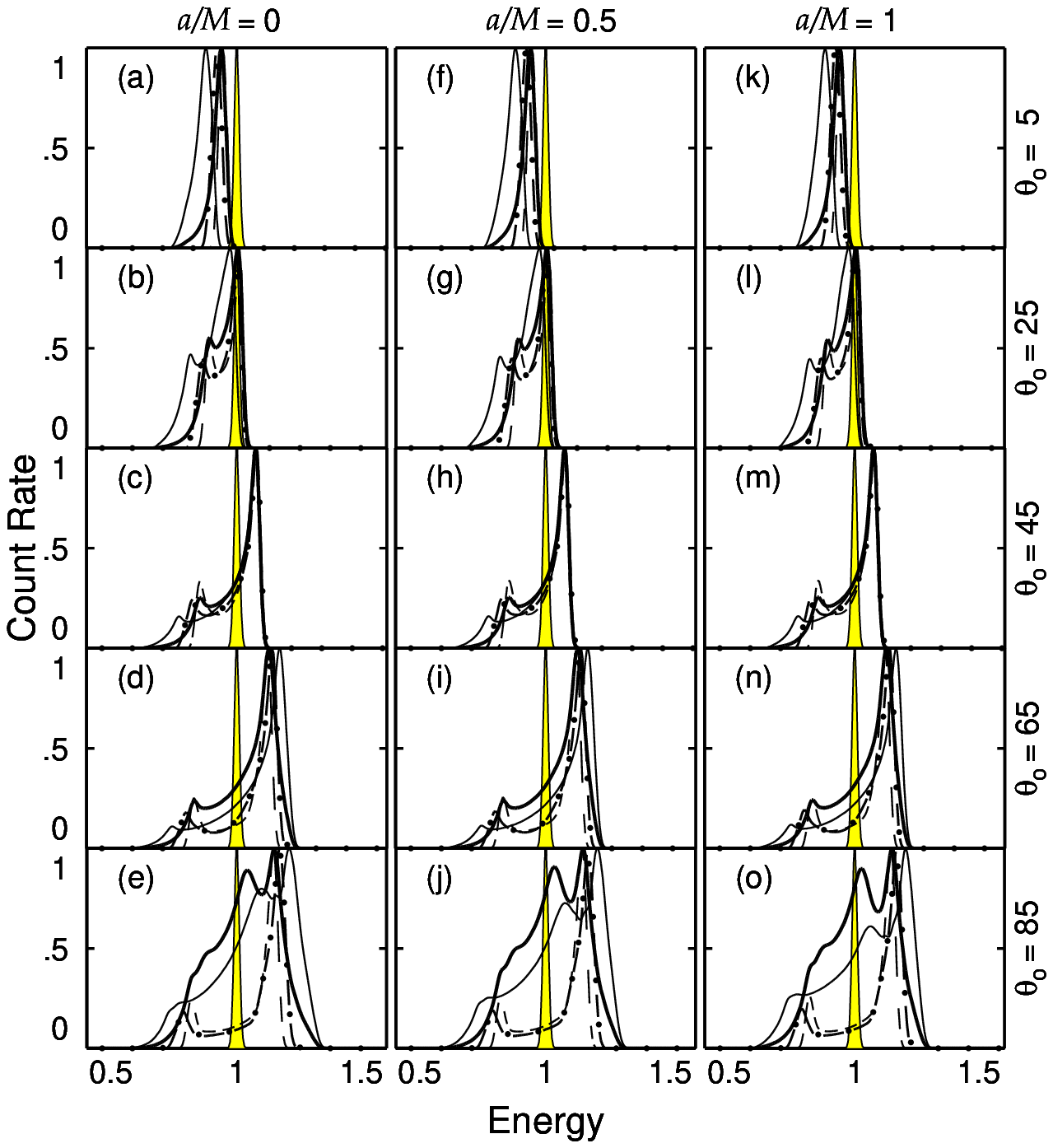}
\caption{As in Fig.~\protect\ref{fig1} but for locally anisotropic radiation.
 \label{fig3}}
\end{figure*}

\begin{figure*}
\centering\leavevmode
\epsfxsize=18cm
\epsfbox[90 187 553 597]{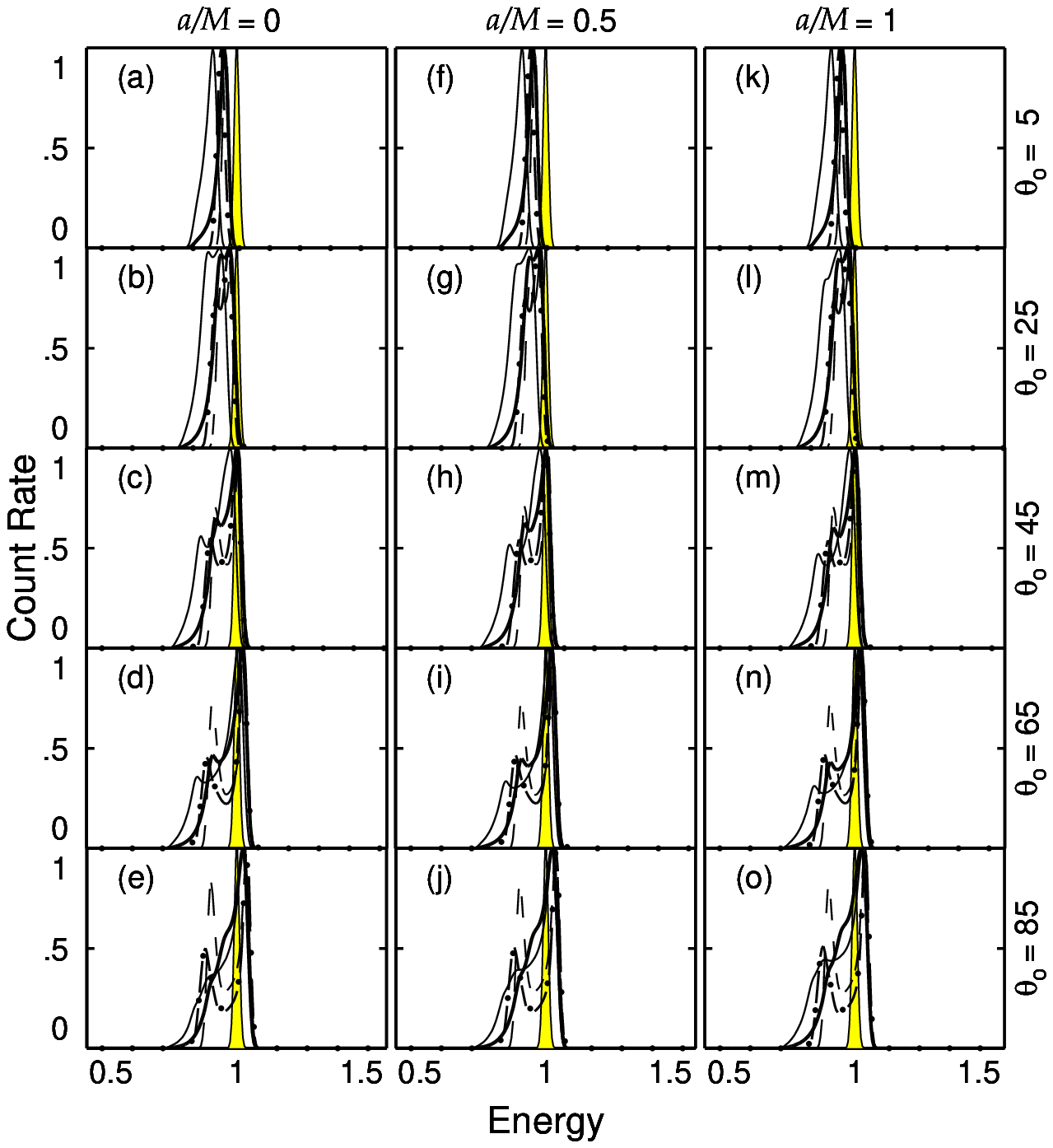}
\caption{As in Fig.~\protect\ref{fig2} but for locally anisotropic radiation.
 \label{fig4}}
\end{figure*}

This Appendix summarizes effects which influence formation of observed
line profiles. We have systematically studied a large set of synthetic
emission-line profiles from an accretion disk around a rotating black
hole which are needed for comparisons with observational data in the
above-described method. Various properties of observed radiation from a
source near a black hole have been discussed by many authors using
different approximations, starting with classical papers by Cunningham
(1976), Laor (1991), and more recently by, e.g., Fanton et al.\ (1997).
Our aim was twofold: (i)~to determine the range of redshift factor
affecting the predicted line profiles under different situations;
(ii)~to see how sensitive the results are to uncertainties in current
models of the disk emission. We considered regions with different radial
distance from the central black hole and took into account also
non-negligible radial motion of gaseous material, different local
profiles of emission lines and the disk shape. Typical results are
illustrated in Figures \ref{fig1}--\ref{fig4}.

The model of the disk is characterized by radiation intensity
$I_{\rm em}$ emitted locally from its surface in the equatorial
plane at frequency $\nu_{\rm em}$,
\begin{equation}
  I_{\rm em}(\nu_{\rm em};r_{\rm em},\theta_{\rm em})=
  F(r_{\rm em})\varphi_1(\nu_{\rm em})\varphi_2(\vartheta),
\end{equation}
where $F(r_{\rm{}em})$ is total radiation flux at the disk surface,
$\varphi_1(\nu_{\rm{}em})$ is the emissivity profile in frequency,
$\varphi_2(\vartheta)$ is the limb-darkening law; $\vartheta$ denotes
the angle between the ray and direction normal to the disk in LDF. We
have classified different models according to following parameters:

\placefigure{fig1}
\placefigure{fig2}
\placefigure{fig3}
\placefigure{fig4}
\placetable{tab2}

\begin{enumerate}
\item{}Local radiation flux, $F(r_{\rm em})$.
  We examined:
  (i)~power-law dependence
      $F(r_{\rm em})\propto{r_{\rm em}}^{-\beta}$
      ($\beta={\rm const}$);
  (ii)~standard Novikov \& Thorne (1973) model.

\item{}Emissivity profile in frequency,
       $\varphi_1(\nu_{\rm em})$.
  We examined two cases:
  (i)~Gaussian profiles
      $\varphi_1(\nu_{\rm em})$ $\propto$
      $\exp[-\varepsilon(\nu_{\rm em}-1)]^2$
      ($\varepsilon={\rm const}$).
      The frequency profile is normalized to maximum at unit
      local frequency [thus the term $(\nu_{\rm em}-1)$].
  (ii)~Asymmetric profiles which account for self-irradiation of the
      disk (this topic was discussed by several authors but effects of
      irradiation of the disk are still an open problem; cf.\ Petrucci
      \& Henri 1997; Dabrowski et al.\ 1997).

\item{}Angular dependence of the local emissivity (limb-darkening
  law), $\varphi_2(\vartheta)$.
  We examined:
  (i)~$\varphi_2(\vartheta)\propto1+\epsilon\mu$; $\mu=\cos\theta$,
      $\epsilon={\rm const}$ ($\epsilon=0$ corresponds to
      locally isotropic emission).
  (ii)~$\varphi_2(\vartheta)\propto\mu\log(1+1/\mu)$ (Ghisellini,
      Haardt, \& Matt 1994).
\end{enumerate}

Ray-tracing of the light trajectories was performed in the Kerr
geometry with corresponding redshift function $g$ being given by
\begin{eqnarray}  \label{gfun}
  g & = &
   \frac{\hat{p}_\alpha\hat{\eta}^\alpha}{p_\alpha\eta^\alpha}
   \;=\;g^{tt}\hat{\eta}_t
   +g^{t\phi}\left(\hat{\eta}_\phi-\xi\hat{\eta}_t\right)
     \nonumber \\
   & &
       -g^{\phi\phi}\xi\hat{\eta}_\phi
       +g^{rr}\hat{\eta}_r\hat{p}_r/p_t\,,
\end{eqnarray}
and
\begin{equation}  \label{angle}
  \cos\vartheta=-\frac{p_{\alpha}n^{\alpha}}
                      {p_{\alpha}\eta^{\alpha}} \; .
\end{equation}
Here, $p^\alpha$ and $\eta^\alpha$ denote respectively the four-momenta
of the photon and of the emitting material in the disk, and analogously
$\hat{p}^\alpha$ and $\hat{\eta}^\alpha$ that of the photon and of
the observer at $r\rightarrow\infty$; $n^{\alpha}$ denotes a unit
space-like vector perpendicular to the disk surface.

The resulting synthetic lines are characterized by observed radiation
flux (count rate) as a function of energy. Predicted profiles are shown
in figures \ref{fig1}--\ref{fig4} (background is subtracted and flux is
normalized to maximum). Characteristics of the line profiles can be
represented in terms of the following parameters:

\begin{enumerate}
\item{}Normalized centroid energy $E^{\rm c}$ (Matt, Perola \&
  Piro 1991) which is determined by energy distribution $N(E)$ of
  the line photons:
\begin{equation}
  E^{\rm c}=N^{-1}\int{}N(E)E{\rm d}E \, , \quad{}
          N=\int{}N(E)\,{\rm d}E \, ;
\end{equation}
  $E^{\rm c}$ characterizes the shift of the line energy (notice
  that $E^{\rm c}$ is not the standard energy centroid which is
  defined through the continuum flux).

\item{}Ratio of line widths (of the observed width to the width in LDF)
  $\delta\sigma$. Here, geometrical line width $\sigma(E^{\rm{}c})$ is
  defined by
\begin{equation}
  \sigma^2(E^{\rm c})=
  {N^{-1}\int{}N(E)(E-E^{\rm c})^2{\rm d}E} \; .
\end{equation}
  Broadening of the line corresponds to $\delta\sigma>1$.

\item{}Normalized (to the unit continuum flux) equivalent width
\begin{equation}
  W=\int{}N(E)E\,{\rm d}E \, .
\end{equation}

\item{}Energy $F^{\rm{c}}$ of the line center, given by
\begin{equation}
  \int_0^{F^{\rm c}}N(E)E\,{\rm d}E=
  \int_{F^{\rm c}}^{\infty}N(E)E\,{\rm d}E \, .
\end{equation}

\item{}Energy of the maximum count rate, $F^{\rm m}$.
\end{enumerate}

In Fig.~\ref{fig1}, the emitted line profile (in the local frame
corotating with the disk material) is indicated by shading; here, it is
the Gaussian line originating in a thin Keplerian disk. Photon energy is
normalized to unity and indicated on the horizontal axis. Locally
emitted radiation flux is isotropic ($\epsilon=0$) and decreases as
$\propto{r^{-1/2}}$. The thick solid curve corresponds to the observed
profile from the whole disk which extends between
$r_{\rm{in}}=6M<r<25M=r_{\rm{out}}$. Contributions from three equally
wide regions between $r_{\rm{in}}$ and $r_{\rm{out}}$ are shown by thin
solid curve (inner part of the disk), dot-dashed curve (central part),
and the dashed curve (outer part). In Fig.~\ref{fig2}, parameters are
the same except for local velocity of the material which has advective
component $v_r=v_\phi=0.4v_{\rm{+}}$; $v_{\rm{+}}$ is the local
corotating Keplerian orbital velocity, $v_r$ and $v_\phi$ are
three-velocity components in the frame corotating with the disk
material. (Numerical value of $v_\phi/v_r$ is model-dependent; here, it
has been chosen for the sake of comparison with Fukue \& Ohna 1997;
Watanabe \& Fukue 1996 suggested an alternative scheme of an advective
corona.) Profiles given in Fig.~\ref{fig3} are constructed as in
Fig.~\ref{fig1} but for locally anisotropic radiation ($\epsilon=2$),
and similarly Fig.~\ref{fig4} illustrates profiles from the disk with
both radial flow and locally anisotropic radiation.

Parameters characterizing geometrical shapes end energy shift of the
profiles from all four figures are summarized in Tab.~2. All
the line profiles in this example have $F^{\rm{c}}_0=E^{\rm{c}}_0=1$,
$W_0=105$ in the local disk frame (subscript ``0''). Much more extensive
set of predicted profiles and corresponding tables are available in the
electronic form from the authors.

\clearpage
\begin{table}
\dummytable\label{tab2}
\end{table}
\clearpage


\clearpage

\clearpage

\clearpage

\clearpage
\pagestyle{empty}

\begin{figure*}
\centering\leavevmode
\epsfxsize=18cm
\epsfbox{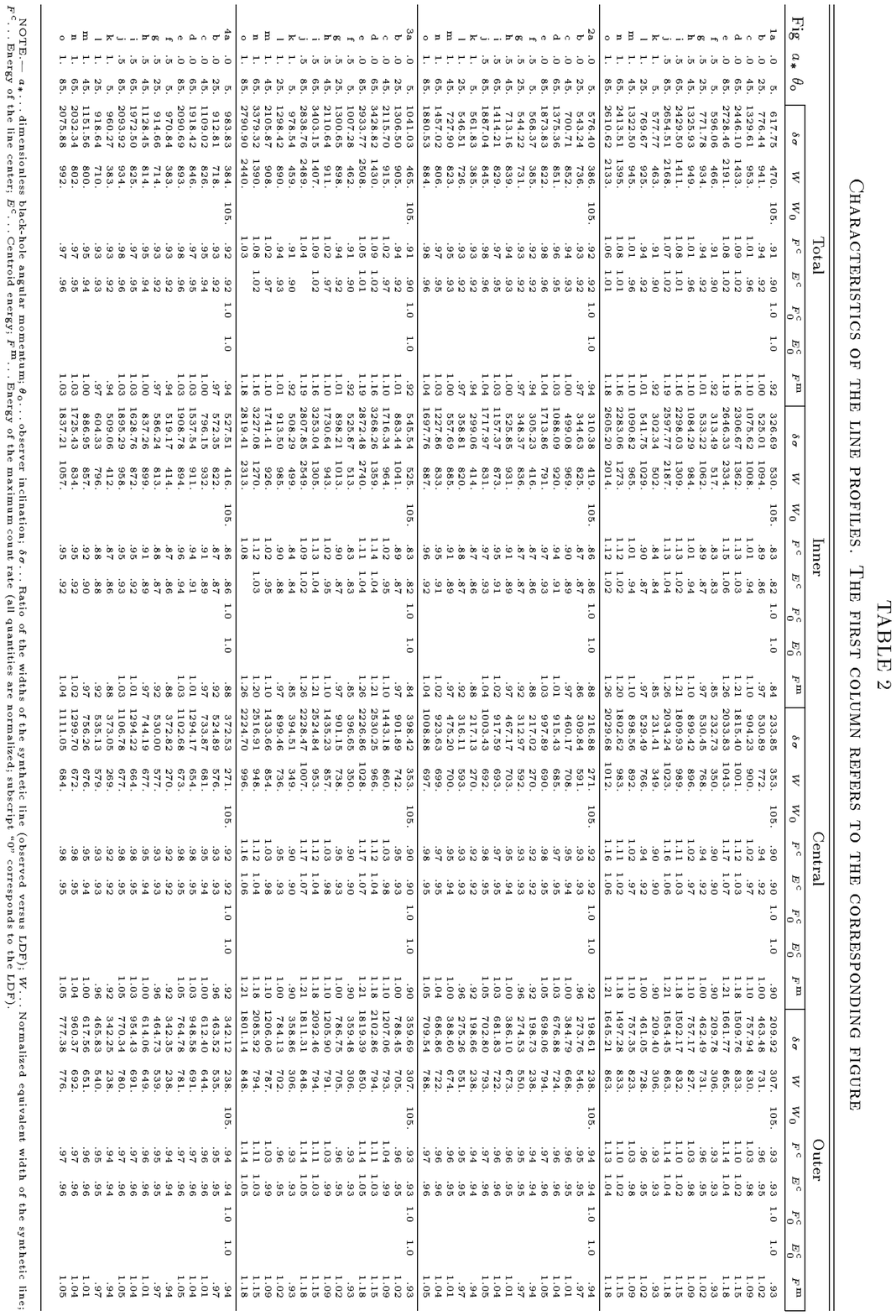}
\end{figure*}
\clearpage


\begin{thebibliography}{99}
\bibitem{A89}
   Asaoka, I.\ 1989,
   \pasj, 41, 763
\bibitem{BHO94}
   Bao, G., Hadrava, P., \& {\O}stgaard, E.~1994,
   \apj, 425, 63
\bibitem{BHO94&96}
   Bao, G., Hadrava, P., \& {\O}stgaard, E.
   1994, \apj, 435, 55, and
   1996, \apj, 464, 684
\bibitem{BHWX97}
   Bao, G., Hadrava, P., Wiita, P.~J., \& Xiong, Y.\ 1997,
   \apj, 487, 142
\bibitem{BPT72}
   Bardeen, J.~M., Press, W.~H., \& Teukolsky, S.~A.\ 1972,
   \apj, 178, 347
\bibitem{BMCG96}
   Barret, D., McClintock, J.~E., \& Grindlay, J.~E.\ 1996,
   \apj, 473, 963
\bibitem{BSNCh96}
   Beekman, G., Shahbaz, T., Naylor, T., \& Charles, P.~A.\ 1996,
   \mnras, 281, L1
\bibitem{BSNChWM97}
   Beekman, G., Shahbaz, T., Naylor, T., \& Charles, P.~A.,
   Wagner, R.~M., \& Martini, P.\ 1997,
   \mnras, 290, 303
\bibitem{BBR84}
   Begelman, M.~C., Blandford, R.~D., \& Rees, M.~J.\ 1984,
   Rev.\ Mod.\ Phys., 56, 255
\bibitem{BvdKLPDMM97}
   Belloni, T., van der Klis, M., Lewin, W.~H.~G., van Paradijs,
   J., Dotani, T., Mitsuda, K., \& Miyamoto, S.\ 1997,
   \aap, 322, 857
\bibitem{BMcK82}
   Blandford, R.~D., \& McKee, C.~F.\ 1982, \apj, 255,~419
\bibitem{BR92}
   Blandford, R.~D., \& Rees, M.~J.\ 1992,
   in Testing the AGN Paradigm (AIP Conf.\ Proc.\ 254),
   eds.\ S.~S.\ Holt, S.~G.\ Neff, \& C.~M.\ Urry
   (New York: Am.\ Inst.\ Phys.), p.~3
\bibitem{BChM97}
   Bromley, B.~C., Chen, K., \& Miller, W.~A.\ 1997,
   \apj, 475,~57
\bibitem{Ch97}
   Charles, P.~A.\ 1997,
   in Proc.\ 18th Texas Conf.\ (Singapore: World Scientific),
   in press
\bibitem{CZCh98}
   Cui, W., Zhang, S.~N., \& Chen,~W.\ 1998,
   \apj, 492,~L53
\bibitem{C75}
   Cunningham, C. T.\ 1976, \apj, 208,~534
\bibitem{DFILR97}
   Dabrowski, Y., Fabian, A.~C., Iwasawa, K., Lasenby, A.~N., \&
   Reynolds, C.~S.\ 1997, \mnras, 288,~L11
\bibitem{deF79}
   de Felice, F.\ 1979, Physics Letters, 69A,~307
\bibitem{deF94}
   de Felice, F. 1994,
   Class.\ Quantum\ Grav., 11,~1283
\bibitem{deFUT92}
   de Felice, F., \& Usseglio-Tomasset, S.\ 1992,
   Gen.\ Rel.\ Grav., 24, 1091
\bibitem{deFUT96}
   de Felice, F., \& Usseglio-Tomasset, S.\ 1996,
   Gen.\ Rel.\ Grav., 28, 179
\bibitem{DFKNP97}
   Diener, P., Frolov, V.~P., Khokhlov, A.~M., Novikov, I.~D., \&
   Pethick, C.~J.\ 1997, \apj, 479,~164
\bibitem{EG97}
   Eckart, A., \& Genzel,~R.\ 1997,
   \mnras, 284,~576
\bibitem{F97}
   Fabian, A.~C.\ 1997,
   Astron.\ Geophys., 38,~10
\bibitem{FNRBOTII95}
   Fabian, A.~C., Nandra, K., Reynolds, C.~S., Brandt, W.~N.,
   Otani, C., Tanaka, Y., Inoue, H., \& Iwasawa, K.\ 1995,
   \mnras, 277, L11
\bibitem{FCdeFC97}
   Fanton, C., Calvani, M., de Felice, F., \& \v{C}ade\v{z}, A.
   1997,
   \pasj, 49, 159
\bibitem{FFJ96}
   Ferrarese, L., Ford, H.~C., \& Jaffe, W.\ 1996,
   \apj, 470, 444
\bibitem{FR76}
   Frank, J., \& Rees, M.~J.\ 1976,
   \mnras, 176, 633
\bibitem{FO97}
   Fukue, J., \& Ohna,~E.\ 1997,
   \pasj, 49, 315
\bibitem{GBO'D97}
   Gallimore, J.~F., Baum, S.~A., \& O'Dea, C.~P.\ 1997,
   \nat, 388, 852
\bibitem{GHM94}
   Ghisellini, G., Haardt, F., \& Matt,~G. 1994,
   \mnras, 267, 743
\bibitem{HMP94}
   Hameury, J.-M., Marck, J.-A., \& Pelat,~D.\ 1994,
   \aap, 287, 795
\bibitem{I94}
   Ipser, J.~R. 1994, \apj, 435,~767
\bibitem{IFBKMRT98}
   Iwasawa,~K., Fabian, A.~C., Brandt, W.~N., Kunieda,~H.,
   Misaki,~K., Reynolds, C.~S., \& Terashima,~Y. 1998,
   \mnras, submitted (astro-ph/9801226)
\bibitem{IFRNOIHBDKMT96}
   Iwasawa, K., Fabian, A.~C., Reynolds, C.~S., Nandra, K.,
   Otani, C., Inoue, H., Hayashida, K., Brandt, W.~N., Dotani,
   T., Kunieda, H., Matsuoka, M., \& Tanaka, Y.\ 1996,
   \mnras, 282, 1038
\bibitem{JFFvdBOC93}
   Jaffe, W., Ford, H.~C., Ferrarese, L., van den Bosch, F., \&
   O'Connell, R.~W.\ 1993,
   \nat, 364, 213
\bibitem{KTF95}
   Kanetake, R., Takeuti, M., \& Fukue, J.\ 1995,
   \mnras, 276, 971
\bibitem{K96}
   Karas, V. 1996, \apj, 470, 743
\bibitem{KV94}
   Karas, V., \& Vokrouhlick\'y, D.\ 1994,
   \apj, 422, 218
\bibitem{KKS97}
   King, A.~R., Kolb, U., \& Szuszkiewicz, E.\ 1997,
   \apj, 488, 89
\bibitem{KR95}
   Kormendy, J., \& Richstone, D.\ 1995,
   \araa, 33, 581
\bibitem{KBADFGGLRT96}
   Kormendy, J., Bender, R., Ajhar, E.~A.,
   Dressler, A., Faber, S.~M., Gebhardt, K., Grillmair, C.,
   Lauer, T.~R., Richstone, D., \& Tremaine, S.~1996b,
   \apj, 473, L91
\bibitem{KBMTGRDFGL97}
   Kormendy, J., Bender, R., Magorrian, J., Tremaine, S.,
   Gebhardt, K., Richstone, D., Dressler, A., Faber, S.~M.,
   Grillmair, C., \& Lauer, T.~R.\ 1997,
   \apj, 482, L139
\bibitem{KBRADFGGLT96}
   Kormendy, J., Bender, R., Richstone,~D., Ajhar, E.~A.,
   Dressler,~A., Faber, S.~M., Gebhardt,~K., Grillmair,~C.,
   Lauer, T.~R., \& Tremaine,~S.\ 1996a,
   \apj, 459,~L57
\bibitem{KHKMEK91}
   Krolik, J.~H., Horne, K., Kallman, T.~R., Malkan, M.~A.,
   Edelson, R.~A., Kriss, G.~A., 1991, \apj, 371, 541
\bibitem{L91}
   Laor, A.\ 1991, \apj, 376, 90
\bibitem{LA97}
   Lasota, J.-P., \& Abramowicz, M.~A.\ 1997,
   Class.\ Quantum Grav., 14, A237
\bibitem{LT18}
   Lense, J., \& Thirring, H.\ 1918, Phys.\ Zeits., 19, 156
\bibitem{LPvdH95}
   Lewin, W.~H.~G., van Paradijs, J., \& van den Heuvel, E.~P.
   J., eds.\ 1995, X-ray Binaries
   (Cambridge: Cambridge Univ.\ Press)
\bibitem{LU97}
   Loeb, A., \& Ulmer, A.\ 1997, \apj, 489, 573
\bibitem{MMACSC97}
   Macchetto, F., Marconi, A., Axon, D.~J., Capetti, A., Sparks,
   W., \& Crane, P.\ 1997, \apj, 489, 579
\bibitem{MLB96}
   Marck, J.~A., Lioure, A., \& Bonazzola, S.\ 1996
   \aap, 306, 666
\bibitem{ML98}
   Markovi\'c, D., \& Lamb, F.~K. 1998, astro-ph/9801075
\bibitem{MM96}
   Martocchia, A., \& Matt, G.\ 1996,
   \mnras, 282, L53
\bibitem{MPPS92}
   Matt, G., Perola, G.~C., Piro, L., \& Stella, L.\ 1992,
   \aap, 257, 63
\bibitem{MO97}
   Merritt, D., \& Oh, S.~P.\ 1997, \aj, 113, 1279
\bibitem{MDZ96}
   Mezger, P.~G., Duschl, W.~J., \& Zylka, R.\ 1996, \aap, 7, 289
\bibitem{MON94}
   Mineshige, S., Ouchi, N.~B., Nishimori, H. 1994, \pasj, 46,~97
\bibitem{MTW}
   Misner, C.~W., Thorne, K.~S., \& Wheeler, J.~A.\ 1973,
   Gravitation (San Francisco: Freeman)
\bibitem{NGMTY97}
   Nandra, K., George, I.~M., Mushotzky, R.~F., Turner, T.~J., \&
   Yaqoob, T.\ 1997, \apj, 476, 70 (I); \apj, 477, 602 (II)
\bibitem{NMYGT97}
   Nandra, K., Mushotzky, R.~F., Yaqoob, T., George, I.~M., \&
   Turner, T.~J.\ 1997, \mnras, 284, L7
\bibitem{NGM97}
   Narayan, R., Garcia, M.~R., \& McClintock, J.~E.\ 1997,
   \apj, 478, L79
\bibitem{NT73}
   Novikov, I.~D., \& Thorne, K.~S.\ 1973, in: Black Holes, eds.\
   C.~DeWitt, B.~DeWitt (New York: Gordon \& Breach), p.~343
\bibitem{OB97}
   Orosz, J.~A., \& Bailyn, C.~D.\ 1997, \apj, 477, 876
\bibitem{PL93}
   Papadakis, I.~E., \& Lawrence, A.\ 1993, \nat, 361, 233
\bibitem{PB98}
   Pariev, V.~I., \& Bromley, B.~C.\ 1998,
   in Proc.\ of the 8th Ann.\ Oct.\ Astrophys.\ Conf.\ in
   Maryland, to appear (astro-ph/9711214)
\bibitem{PMCChK96}
   Pavlenko, E.~P., Martin, A.~C., Casares, J., Charles, P.~A.,
   \& Ketsaris, N.~A.\ 1996, \mnras, 281,~1094
\bibitem{PH97}
   Petrucci, P.~O., Henri, G.\ 1997, \aap, 326,~99
\bibitem{PR94}
   Podsiadlowski, P., \& Rees, M.~J.\ 1994,
   in The Evolution of X-ray Binaries, AIP Conf.~Proc.\ 308,
   eds.\ S.~S.\ Holt \& C.~S.\ Day
   (New York: AIP Press), p.~71
\bibitem{R84}
   Rees, M.~J.\ 1984, \araa, 22,~471
\bibitem{R98}
   Rees, M.~J.\ 1998, in Black Holes and Relativistic Stars,
   ed.\ R.~M. Wald (Chicago: Univ.\ Chicago Press), in press
\bibitem{RB97}
   Reynolds, C.~S., \& Begelman, M.~C.\ 1997, \apj, 488,~109
\bibitem{RB98}
   Rybicki, G.~B., \& Bromley, B.~C.\ 1998,
   \apj, submitted (astro-ph/9711104)
\bibitem{SdeF97}
   Semer\'ak, O., \& de Felice, F.\ 1997,
   Class.\ Quantum Grav., 14,~2381
\bibitem{Sl98}
   \v{S}lechta, M.\ 1998,
   unpublished PhD Thesis (Prague: Charles University)
\bibitem{SC77}
   Stark, R.~F., \& Connors, P.~A. 1977,
   \nat, 266,~429
\bibitem{SCS97}
   Steiman-Cameron, T.~Y., \& Scargle, J.~D.\ 1997,
   \apj, 487,~396
\bibitem{St90}
   Stella, L.\ 1990,
   \nat, 344,~747
\bibitem{SS97}
   Stothers, R.~B., \& Sillanp\"a\"a,~A.\ 1997,
   \apj, 475,~L13
\bibitem{SWLV97}
   Sundelius,~B., Wahde,~M., Lehto, H.~J., \& Valtonen, M.~J. 1997,
   \apj, 484,~180
\bibitem{SyC95}
   Syer, D., \& Clarke, C.~J.\ 1995,
   \mnras, 277, 758
\bibitem{SCR92}
   Syer, D., Clarke, C.~J., \& Rees, M.~J.\ 1991,
   \mnras, 250,~505
\bibitem{TBIST96}
   Tagliaferri, G., Bao, G., Israel, G.~L., Stella, L., \&
   Treves, A.\ 1996,
   \apj, 465,~181
\bibitem{TNFIODHIKKMM95}
   Tanaka,~Y., Nandra,~K., Fabian, A.~C., Inoue,~H., Otani,~C.,
   Dotani,~T., Hayashida,~K., Iwasawa,~K., Kii,~T., Kunieda,~H.,
   Makino,~F., \& Matsuoka,~M.\ 1995,
   \nat, 375,~659
\bibitem{VL97}
   Valtonen, M.~J., \& Lehto, H.~J.\ 1997,
   \apj, 481,~L5
\bibitem{vdBdZ96}
   van den Bosch, F.~C., \& de Zeeuw, P.~T.\ 1996,
   \mnras, 283,~381
\bibitem{vdMZR97}
   van der Marel, R.~P., de Zeeuw, P.~T., \& Rix, H.-W.\ 1997,
   \apj, 488,~119
\bibitem{VRST98}
   Villata,~M., Raiteri, C.~M., Sillanp\"a\"a,~A., \& Takalo, L.~O.
   1998, \mnras, 293,~L13
\bibitem{VK93}
   Vokrouhlick\'y, D., \& Karas, V.\ 1993,
   \mnras, 265, 365
\bibitem{VK98}
   Vokrouhlick\'y, D., \& Karas,~V.\ 1998,
   \mnras, 293,~L1
\bibitem{W72}
   Wilkins, D.~C.\ 1972,
   \prd, 5, 814
\bibitem{YSTGN96}
   Yaqoob, T., Serlemitsos, P.~J., Turner, T.~J., George, I.~M.,
   \& Nandra, K.\ 1996,
   \apj, 470, L27
\bibitem{Z94}
   Zakharov, A.~F.\ 1994,
   \mnras, 269, 283
\bibitem{ZCCh97}
   Zhang, S.~N., Cui, W., \& Chen, W.\ 1997,
   \apj, 482, L155
\bibitem{ZSC94}
   Zurek, W.~H., Siemiginowska, A., \& Colgate, S.~A.\ 1994,
   \apj, 434, 46
\end{thebibliography}
\end{document}